\documentclass[12pt, A4]{article}

\usepackage[utf8]{inputenc}

\usepackage{graphicx}
\usepackage{caption}
\usepackage{subcaption}
\usepackage{booktabs}
\usepackage{multirow}
\usepackage{rotating}
\usepackage{diagbox}

\usepackage{diagbox}

\usepackage[breaklinks=true]{hyperref}
\usepackage{amsmath}
\usepackage{xcolor}

\usepackage{times}
\usepackage{natbib}
\usepackage{hyperref}
\usepackage{amsfonts}

\usepackage{graphicx}
\usepackage{rotating}
\usepackage{diagbox}

\usepackage{slashbox}
\usepackage{bm}

\usepackage{setspace}
 \usepackage[left=2.5cm,right=2.5cm,top=2.5cm,bottom=2.5cm]{geometry}

\def\bSig\mathbf{\Sigma}

 \newcommand{\ind}{\perp\!\!\!\!\perp} 
\makeatother

\newtheorem{lemma}{Lemma}
\newtheorem{assumption}{Assumption}
\newtheorem{corollary}{Corollary}
\newtheorem{appendixassumption}{Assumption}

\newtheorem{theorem}{Theorem}
  \title{Valid causal inference with unobserved confounding in high-dimensional settings}

\author{Niloofar Moosavi, Tetiana Gorbach, Xavier de Luna \\
\\
{\it Department of Statistics, USBE, Umeå University, Umeå, Sweden} \\
{\it  mousavi.n0@gmail.com, tetiana.gorbach@umu.se, xavier.deluna@umu.se}}

\date{}

\begin{document}
\maketitle

\begin{abstract}{Various methods have recently been proposed to estimate causal effects with confidence intervals that are uniformly valid over a set of data generating processes when high-dimensional nuisance models are estimated by post-model-selection or machine learning estimators. These methods typically require that all the confounders are observed to ensure identification of the effects. We contribute by showing how valid semiparametric inference can be obtained in the presence of unobserved confounders and high-dimensional nuisance models. We propose uncertainty intervals which allow for unobserved confounding, and show that the resulting inference is valid when the amount of unobserved confounding is small relative to the sample size; the latter is formalized in terms of convergence rates. Simulation experiments illustrate the finite sample properties of the proposed intervals and investigate an alternative procedure that improves the empirical coverage of the intervals when the amount of unobserved confounding is large. Finally, a case study on the effect of smoking during pregnancy on birth weight is used to illustrate the use of the methods introduced to perform a sensitivity analysis to unobserved confounding.}
\end{abstract}

\section{Introduction}\label{s:intro}
In contrast to randomized experiments, observational studies are prone to the presence of confounding variables which are not balanced among treated and control individuals \citep{fisher1958cigarettes}. In such studies, all confounders are often assumed to be observed for identifiability of the causal parameter of interest \citep{rubin1974estimating, DR:90}. Efforts to make this assumption plausible and the use of flexible nuisance models often yield a high-dimensional setting, where the number of nuisance parameters to be fitted may be at least of the order of the sample size. 
Nonetheless, some confounders may still be unobserved. Therefore, well-conducted observational studies should investigate the sensitivity of the inference to the assumption of no unmeasured confounding \citep{cornfield1959smoking}.

In this paper, we study sensitivity analysis for semiparametric inference on the average causal effect of a binary treatment in high-dimensional observational studies. In this context, under certain conditions, augmented inverse probability weighting (\textsc{AIPW}) \citep*{robins1994estimation} and targeted learning \citep{vdLR:11} estimators yield uniformly valid inferences even when nuisance models are fitted with flexible machine learning algorithms; see, e.g., \linebreak[4]\citet{farrell2015robust,van2010collaborative, chernozhukov2018double} and \citet*{belloni2014inference, moosavi2021costs} for a recent review. Roughly, uniformly valid inference means that the finite sample behavior of the estimator can be well approximated by its asymptotic distribution, even if preliminary analysis, such as variable selection, has been performed on the nuisance models prior to final estimation.  \cite{farrell2015robust} studied uniformly valid inference associated with the \textsc{AIPW} estimator when flexible estimators of nuisance models are weakly consistent and fulfill multiplicative rate conditions (See Assumption \ref{assum:rates} below). 
Here, we build on these results to allow for unobserved confounding and study the uniform validity of the resulting inference. 
For this, we first specify the confounding bias of the AIPW estimator of the causal effect as a function of unobserved confounding. We then propose uncertainty intervals for the causal effect which account for such bias \citep{vansteelandt2006ignorance,genback2019causal}. 
  Our suggested inference on the causal effect using the uncertainty intervals ignores the finite sample bias and randomness in the estimation of the confounding bias. It provides uniformly valid inference given assumptions on the amount of unobserved confounding relative to the sample size; the latter is  formalized  in  terms  of  convergence  rate.  We provide simulations to illustrate the relevance of the theory for finite samples.   

This work contributes to the sensitivity analysis literature that originated in \citet{cornfield1959smoking} and has been studied further by many others \citep*[to cite only a few:][]{rosenbaum1987sensitivity,  ding2016sensitivity, franks2019flexible,bonvini2019sensitivity, zhang2019semiparametric, zhao2019sensitivity,erin:2023}. More specifically, we contribute to research which considers the sensitivity parameter to be the correlation between the potential outcomes and the treatment assignment given the observed covariates induced by unmeasured confounders \citep*{copas1997inference, imai2010general, genback2019causal}. Our contribution is to consider post-model-selection inference, including high-dimensional situations, and inference following machine learning fits of the nuisance models. In this respect, our work is most related to \citet{scharfstein2021semiparametric}, which also considers inference on a causal parameter based on flexible estimation of nuisance models and allows for unobserved confounders via sensitivity analysis. Their approach differs in how the confounding bias is parameterized. We contrast the methods using a case study in Section 4. 

The rest of the paper is organised as follows. Section 2 describes the context and states our result on the uniform validity of the inference for a post-model-selection estimator of a causal parameter, for a given amount of unobserved confounding described by a correlation parameter. In practice, this correlation is unknown and a plausible range of correlation values may be considered as a sensitivity analysis. In Section 3, simulations are provided to illustrate the relevance of the theory for finite samples, both in low- and high-dimensional settings. Section 4 presents a case study of the effect of maternal smoking on birth weight. This illustrates the use of the herein proposed methods implemented in the \texttt{R}-package \texttt{hdim.ui} (\url{https://github.com/stat4reg/hdim.ui}). Section 5 concludes the paper.


\section{Theory and method}
 We aim to study the causal effect of a binary treatment $T$ on an outcome of interest $Y.$ Let $Y(t)$, $t \in \{0,1\}$, be the potential outcomes that would have been observed under a corresponding treatment level, and assume the observed outcome is $Y = TY(1) + (1-T)Y(0)$, for all units in the study. 
 The average causal effect of the treatment is defined as $E\left(Y\left(1\right) - Y\left(0\right)\right)$. Without loss of generality, we focus on the parameter $\tau = E\left(Y\left(1\right)\right)$; see the Appendix for other parameters of interest. We denote the vector of pre-treatment covariates and their transformations by $X = (1, X^{(1)}, \ldots, X^{(p-1)})$. For simplicity, we call this the set of covariates. The dimension $p$ of this vector is considered to increase with the sample size $n$.
    
    Consider $O_{i,n}=\left\{\left(X_{i,n}, Y_{i,n}, T_{i,n}\right) \right\}_{i=1}^n$ to be an i.i.d. sample that follows a distribution $P_n$ which is allowed to vary with the sample size $n$. We drop the subscript $n,$ which implies dependence on $n,$ when it is clear from the context. Moreover, let $n_t$ denote the number of individuals with observed treatment $T=1$. 
    We use the notation $E_n[W_i] = n^{-1}\Sigma_{i=1}^n W_i$,  $E_n[W_i]^q = (n^{-1}\Sigma_{i=1}^n W_i )^q$  and $E_{n_t}[W_i] = n_t^{-1} \Sigma_{i=1}^{n} T_i W_i$. 

We consider the augmented inverse propensity weighting estimator (\textsc{AIPW} ; See \citet[ ][]{robins1994estimation} and \citet*{scharfstein1999rejoinder}):
\begin{equation}\label{DR}
   \hat{\tau}_{\textsc{AIPW}} = E_n\left[\dfrac{T_i\left(Y_i-\hat{m}\left(X_i\right)\right)}{\hat{e}\left(X_i\right)} + \hat{m}\left(X_i\right)\right],
\end{equation}
where $\hat{m}(X)$ and $\hat{e}(X)$ are arbitrary estimators of $m(X)=E\left(Y\left(1\right)\mid  X,T=1\right)$ and ${e}(X) = E(T\mid  X)$, respectively. Let us consider the following assumptions. 

\begin{assumption}\label{assum:overlap}
$e(X)>0$.
\end{assumption}

\begin{assumption}\label{assum:rates} Assume $\hat{e}(X)$ is constant conditional on the observed value of $\{X_i,T_i\}_{i=1}^n$ \citep[assumption of "no additional randomness",][ a revision of \citeauthor{farrell2015robust}, \citeyear{farrell2015robust}]{farrell2018robust} and
\begin{enumerate}
    \item[a.] $E_n \left[\left(\hat{e}\left(X_i\right) - {e}\left(X_i\right)\right)^2\right] = o_P(1)$ and $E_n\left[\left(\hat{m}\left(X_i\right) - {m}\left(X_i\right)\right)^2\right] = o_P(1)$.
    \item[b.] $E_n \left[\left(\hat{m}\left(X_i\right) - {m}\left(X_i\right)\right)^2\right]^{1/2}E_n \left[\left(\hat{e}\left(X_i\right) - {e}\left(X_i\right)\right)^2\right]^{1/2} = o_P(n^{-1/2})$.
    \item[c.] $E_n \left[\left(\hat{m}\left(X_i\right) - {m}\left(X_i\right)\right)\left(1 - T_i/{e}\left(X_i\right)\right)\right] = o_P(n^{-1/2})$.
\end{enumerate}
\end{assumption}

The following theorem gives the asymptotic behaviour of the \textsc{AIPW} estimator and its asymptotic target parameter. The theorem generalizes Theorem 3 in \citet{farrell2018robust} by avoiding the assumption $E\left(Y \left(1 \right)\mid X = x,T=1\right) = E\left(Y \left(1 \right)\mid X = x\right)$, i.e., allowing for unobserved confounders.

\begin{theorem}\label{theorem:asym_linearity_AIPW}
Let Assumptions \ref{assum:overlap}, \ref{assum:rates}, and Assumption \ref{DGPfarrellassumptions} in the Appendix hold. The \textsc{AIPW} estimator \eqref{DR} is asymptotically linear as follows:
\begin{equation*}
    \begin{aligned}
\sqrt{n} \left(\hat{\tau}_{\textsc{AIPW}} - \tau^- \right) = \Sigma_{i=1}^n \Psi_i/\sqrt{n} + o_p(1),
    \end{aligned}
\end{equation*}
 where $\tau^-  = E\left( {m}\left(X\right)\right)$ and  $\Psi_i= \dfrac{T_i\left(Y_i-{m}\left(X_i\right)\right)}{{e}(X_i)} + {m}(X_i) - E\left( {m}\left(X_i\right)\right)$.  
\end{theorem} 

Following \cite{genback2019causal}, we study the consequences of the presence of unobserved confounders of the treatment-outcome relationship on inference about $\tau$ by postulating a sensitivity model:

\begin{assumption}\label{assum:sensitivity_model}
Let $Y(1) = E\left(Y\left(1\right)\mid  X\right) + \xi$, $T^* =g(X) + \eta,$ $T = \boldsymbol{I}(T^* > 0),$
where $\eta\sim \mathcal{N}(0,1)$ (i.e., a probit link is used), $\eta \ind X,$ and $\boldsymbol{I}$ represents an indicator function. Moreover, let $\xi = \rho \sigma \eta + \epsilon$, where $\rho= \text{corr} ( \xi, \eta)$, $\sigma^2= \text{var}(\xi) < \infty$ with $\epsilon$ satisfying $\epsilon \ind (X,\eta)$ and $E(\epsilon)=0$.
 \end{assumption}
 
The case $\rho=0$ corresponds to the inclusion of all confounders in $X$. Assumption \ref{assum:sensitivity_model}, therefore, considers a departure from the commonly used identification assumption $E\big(Y \left(1 \right)\mid X = x,T=1\big) = E\big(Y \left(1 \right)\mid X = x\big)$ (no unobserved confounders) in order to perform a sensitivity analysis, and is thus not a restriction of the semiparametric inference performed under the condition of no unobserved confounders. Instead, Assumption \ref{assum:sensitivity_model} allows us to consider a potential amount of unobserved confounding through an interpretable parameter $\rho$, as well as to compute the size of the confounding bias of the \textsc{AIPW} estimator with the following result.

\begin{theorem}\label{theorem:asym_conf_bias_expression}
Suppose that for some $g^0(X)$ and $m^0(X)$ we have \newline $E_{n}\left[|\hat{g}\left(X_{i}\right)-g^0\left(X_{i}\right)|\right]=o_P(1)$ and $E_{n}\left[\left(\hat{m}\left(X_{i}\right)-m^0\left(X_{i}\right)\right)^{2}\right]=o_P(1)$. Assume \ref{assum:overlap} and  \ref{assum:sensitivity_model}. Furthermore, assume \ref{DGPfarrellassumptions} holds for  $m^0(X)$. If $g^0(X)=g(X)$ or $m^0(X)=m(X)$, the (asymptotic) confounding bias of the \textsc{AIPW} estimator \eqref{DR}, which we denote by $b = \tau^- - \tau$, is equal to 
\begin{equation*}
b  =  \rho \sigma E\left(\lambda\left(g\left(X\right)\right)\right),
\end{equation*}
where $\lambda(\cdot)= \phi(\cdot)/\Phi(\cdot),$ and $\phi$ and $\Phi$ are the probability density function and the cumulative distribution function of the standard normal distribution, respectively.
\end{theorem}

Note that under the conditions of \autoref{theorem:asym_linearity_AIPW}, the AIPW estimator \eqref{DR} is a semiparametric efficient estimator of $\tau^-$\citep[see][Example 5]{hines2021demystifying}. However, if confounders of the treatment-outcome relationship are not observed (not included in $X$), $\rho$ and consequently $b$ would be nonzero and hence the AIPW estimate would not be a consistent estimate of $\tau$. It can be shown that for given values of $\rho$ and $\sigma$, the efficient influence function of the bias term has the form $\rho \sigma \lambda\left(g\left(X)\right)\right) - b$. This motivates us to estimate the confounding bias $b$ empirically by plugging in the estimated $g(X_i)$ values.  For a more realistic scenario, the parameter $\sigma$ must also be estimated. First, we use $ \hat{b} = \rho \hat{\sigma} E_n\left[ \lambda\left(\hat{g}\left(X\right)\right)\right]$ as an estimator of the parameter $b$,  where $\hat{\sigma} = E_{n_t}\left[\left(Y\left(1\right)_i - \hat{m}\left(X_i\right)\right)^2\right]^{1/2}$. The estimator of the variance,
$\hat{\sigma}^2$, is biased due to both high dimensionality/variable selection and unobserved confounders. Below, we give the asymptotic properties of $\hat{\tau}_{\textsc{AIPW}} - \hat{b}$ as an estimator of $\tau$, assuming approximate sparsity and known $\rho.$


\begin{theorem}\label{theorem:var_notcorrected_inference}
Let Assumptions \ref{assum:overlap}--\ref{assum:sensitivity_model} hold. Moreover, let Assumptions \ref{DGPfarrellassumptions} and \ref{extraassumptions} hold and assume
$\sqrt{n}\rho^3 = o(1)$,
 $\sqrt{n}\rho E_n\left[\hat{g}\left(X_i\right) - g\left(X_i\right)\right]  =   o_p (1)$ and
  $\sqrt{n}\rho E_{n_t}\left[\left( {m}(X_i)-\hat{m}\left(X_i\right)\right)^2\right]^{1/2} =o_P(1)$.
 We have:
    \begin{itemize}
        \item [{a.}] $\sqrt{n} \left(\left(\hat{\tau}_{\textsc{AIPW}} - \hat{b}\right) - \tau \right) = \Sigma_{i=1}^n \Psi_i/\sqrt{n} + o_p(1),$
        \item [{b.}] $ V^{-1/2}\sqrt{n} \left(\left(\hat{\tau}_{\textsc{AIPW}} - \hat{b}\right) - \tau\right) \rightarrow_d \mathcal{N}(0,1)$,
        \item [{c.}] $\hat{V} - V = o_P(1)$,
    \end{itemize}
where $V= E(\Psi_i^2),$ and $\hat{V} = E_n\left[\dfrac{T_i \left(Y_i - \hat{m}\left(X_i\right)\right)^2}{\hat{e}\left(X_i\right)^2}\right]+ E_n \left[\left(\hat{m}\left(X_i\right) - \hat{\tau}_{\textsc{AIPW}}\right)^2\right]$.
\end{theorem}
As a corollary, the following 95\% confidence interval for $\tau$ given $\rho$ is uniformly valid. 
\begin{corollary}\label{corollary:conf_interval} For each $n$, let ${\mathcal{P}}_n$
be the set of distributions obeying the assumptions of Theorem \ref{theorem:var_notcorrected_inference}. Then, we have:
\begin{equation*}
    \begin{aligned}
\sup_{P \in {\mathcal{P}}_n} \left| {\text{pr}}_P 
\left(\tau \in \left\{\left(\hat{\tau}_{\textsc{AIPW}} - \hat{b}\right) \pm c_\alpha \sqrt{\hat{V}/n}\right\}\right) - \left(1-\alpha\right)\right|  \rightarrow 0,
    \end{aligned}
\end{equation*}
where $c_\alpha = \Phi^{-1}(1 - \alpha/2)$.

\textbf{Proof.}
The corollary follows from Theorem \ref{theorem:var_notcorrected_inference}  \citep[see][Corollary 2]{farrell2015robust}.
\end{corollary}

The proof of Theorems \ref{theorem:asym_linearity_AIPW}--\ref{theorem:var_notcorrected_inference} 
can be found in Appendix. The assumptions on $\rho$ imply that the correlation between the error terms $\xi$ and $\eta$ has to converge to zero as the sample size grows, even if the nuisance models are estimated parametrically. In other words, the amount of unobserved confounding needs to be small relative to the sample size. In practice, $\rho$ is not known and a sensitivity analysis is obtained by considering the 95\% uncertainty interval constructed as the union of all 95\% confidence intervals obtained by varying $\rho\in (\rho_{min},\rho_{max})$, a plausible interval for $\rho$. If the latter interval contains the true value of $\rho$, then this uncertainty interval covers the parameter $\tau$ with at least 95\% probability \citep[][Corollary 4.1.1]{gorbach2018inference}. When no parametric models are assumed and/or the number of covariates is greater than the number of observations, lasso regression \citep{tibshirani1996regression} can be used in the first step to select low-dimensional sets of variables for fitting linear and probit nuisance models. In other words, a linear (in the parameters) regression can be fitted using variables selected by a preliminary lasso regression. A probit regression for the treatment can be fitted  using variables selected by a preliminary probit-lasso regression. 
When it comes to the rate conditions in Assumption \ref{assum:rates}, one can use a post-selection linear model fit for the outcome under common sparsity assumptions on the true data generating process \citep[for post-lasso regression, see][Corollary 5]{farrell2015robust}. We are unaware of any similar result for an estimator in a sparse probit model. However, we investigate the performance of a post-lasso probit regression in the simulation section.

As noted above, the estimator of $\sigma$ used to obtain $\hat b$ in Theorem \ref{theorem:var_notcorrected_inference} is biased. We, therefore, propose to use instead a corrected estimator of $\sigma$. Given the true value of the sensitivity parameter, this corrected estimator of variance gives us a consistent estimate of the causal parameter.

\begin{theorem}
\label{theorem:var_correction_consistency} Let $E_{n}\left[|\hat{g}\left(X_{i}\right)-g\left(X_{i}\right)|\right]=o_P(1)$ and $E_{n}\left[\left(\hat{m}\left(X_{i}\right)-m\left(X_{i}\right)\right)^{2}\right]=o_P(1)$.
Assume, further, that assumptions \ref{assum:overlap}, \ref{assum:sensitivity_model}, \ref{DGPfarrellassumptions} and \ref{appendix.assum} hold. Let
\begin{equation*}
    \hat{\sigma}^{2}_{\text{c}}= E_{n_t} [(Y(1)_i - \hat{m}(X_i))^2]/(1 - \rho^2 E_{n_t}[\hat{g}(X_i) \lambda(\hat{g}(X_i))] - \rho^2 E_{n_t}[ \lambda^2(\hat{g}(X_i))]).
\end{equation*}
Then, we have
\begin{flalign*}
    \hat{\sigma}^{2}_{\text{c}} \xrightarrow[]{p}  {\sigma}^{2}, \\
    \hat{\tau}_\text{AIPW} - \hat{b}_{\text{c}} \xrightarrow[]{p} \tau,
\end{flalign*}
where $ \hat{b}_{\text{c}} = \rho \hat{\sigma}_{\text{c}} E_n[\lambda(\hat{g}(X_i))]$.
\end{theorem}

Furthermore, both Theorem \ref{theorem:var_notcorrected_inference} and Corollary \ref{corollary:conf_interval} hold under weaker regularity conditions when using $\hat{\sigma}_{\text{c}}$ instead of $\sigma$.

\begin{theorem}\label{theorem:var_corrected_inference}
    Let Assumptions \ref{assum:overlap}--\ref{assum:sensitivity_model} hold. Moreover, let Assumptions \ref{DGPfarrellassumptions} and \ref{extraassumptionsalternatrive} hold and assume
$\rho = o(1)$,
 $\sqrt{n}\rho E_n\left[\hat{g}\left(X_i\right) - g\left(X_i\right)\right]  =   o_p (1)$ and
  $\sqrt{n}\rho E_{n_t}\left[\left( {m}(X_i)-\hat{m}\left(X_i\right)\right)^2\right]^{1/2} =o_P(1)$. Then, results in Theorem \ref{theorem:var_notcorrected_inference} and Corollary \ref{corollary:conf_interval} hold for the corrected estimator $\hat{\tau}_\text{AIPW} - \hat{b}_\text{c}$.
\end{theorem}

Note that the condition on unobserved confounding $\sqrt{n}\rho^3 = o(1)$ in Theorem \ref{theorem:var_notcorrected_inference} is replaced with the weaker condition $\rho = o(1)$ in the above theorem. 
This improvment is illustrated in the simulation study below, where the finite sample performance of both estimators is studied.

In the following sections, we use the \texttt{R} package \texttt{hdim.ui} \linebreak[4](\url{https://github.com/stat4reg/hdim.ui}), where the estimators above have been implemented and can be used to obtain uncertainty intervals and perform sensitivity analysis in high-dimensional situations. The package builds on code from the \texttt{ui} package \citep{genback2019causal} (\url{https://cran.r-project.org/web/packages/ui/index.html}).

\section{Simulation study}
 The simulation study in this section aims to illustrate the finite sample behavior of the proposed estimators. Here we consider a high-dimensional setting, while the results for low-dimensional settings are reported in the Supplementary Material.  
Data was generated according to the model in Assumption \ref{assum:sensitivity_model}. All the covariates are generated to be independently and  normally distributed with mean $0$ and variance $1$. The error terms are generated using $(\eta,\xi) \sim \mathcal{N}(0, \Sigma)$ with $\Sigma = \begin{bmatrix}
1 & \rho \\
\rho & 1 
\end{bmatrix}$. 
We let $p=n$ and simulate outcome and treatment as:
\begin{equation*}    
\begin{aligned}
Y(1) &= 2+ X\beta - \rho \lambda (X \gamma) + \xi,\hskip 1cm T^* = X \gamma + \eta,\\
\beta &= 0.6(1,1/2,1/3,1/4,1/5,1,1/2,1/3,1/4,1/5,0,\ldots,0)',
\\
\gamma &= 0.3(1,1/2,1/3,1/4,1/5,1,1,1,1,1,0,\ldots,0)',
    \end{aligned}
\end{equation*}
where both models include covariates that are weakly associated with the dependent variable. Such covariates are likely to be missed in a variable selection step.

In order to investigate the performance of the inference under variable selection, we use $\hat{\tau}_{\textsc{AIPW}}^{\text{refit}} - \hat{b}^{\text{refit}}$ and $\hat{\tau}_{\textsc{AIPW}}^{\text{refit}} - \hat{b}^{\text{refit}}_{\text{c}}.$  In these estimators, $\hat{m}(X)$ is found by refitting the linear regression using variables selected by preliminary linear-lasso regression and $\hat{g}(X)$ is found by refitting the probit model using variables selected by preliminary  probit-lasso regression. The linear-lasso regression is implemented using the \texttt{hdm} package in R \citep*{R, chernozhukov2016hdm}.
The probit-lasso is implemented using the \texttt{glmnet} package \citep{friedman2021package}, where the regularization parameter is found by cross-validation.
Then, 95\% confidence intervals are constructed for a range of values of the  sensitivity parameter $\rho$ according to Corollary \ref{corollary:conf_interval} using the influence-curve based standard error estimator. If empirical coverages are close to nominal, then the corresponding uncertainty intervals will be conservative by construction, as mentioned above \citep[see][]{gorbach2018inference,genback2019causal}. We are particularly interested in investigating coverages of confidence intervals when varying  $\rho$ as sample size increases ($n=500,1000,1500$), since Theorem \ref{theorem:var_notcorrected_inference} requires $\sqrt n\rho^3=o(1)$ for valid inference. The results of the simulation study are based on $500$ Monte-Carlo replications.  

Table \ref{table:results_highdimensional} reports empirical coverages of the 95\% confidence intervals for $\tau$ using $\hat{\tau}_{\textsc{AIPW}}^{\text{refit}} - b^*$, 
$\hat{\tau}_{\textsc{AIPW}}^{\text{refit}} - \hat{b}^{\text{refit}}$ and $\hat{\tau}_{\textsc{AIPW}}^{\text{refit}} - \hat{b}^{\text{refit}}_{\text{c}}$. Here, $b^*$ is an approximation of the confounding bias $b$ using the true values of $\sigma $ and $\rho$ and the Monte Carlo estimate of $E\left(\lambda\left(g\left(X\right)\right)\right).$

\begin{table}[t]
\centering
    \caption{Empirical coverages of 95\% confidence intervals for $\tau$}

\label{table:results_highdimensional}
    \begin{tabular*}{0.97\textwidth}{r|rrrrrrrrrrrrrr}
  \scriptsize \normalfont{Estimator}&  \multicolumn{4}{c}{$\hat{\tau}_{\normalfont{AIPW}}^{\normalfont{refit}} - b^*$} &&   \multicolumn{4}{c} {$\hat{\tau}_{\normalfont{AIPW}}^{\normalfont{refit}} - \hat{b}^{\normalfont{refit}}$}&& \multicolumn{4}{c} {$\hat{\tau}_{\normalfont{AIPW}}^{\normalfont{refit}} - \hat{b}^{\normalfont{refit}}_{\normalfont{c}}$}\\ \hline
 $n$ \textbackslash $\rho$
 & $0.8$ & $0.6$ &$0.4$ & $0.2$ &
 & $0.8$ & $0.6$ &$0.4$ & $0.2$ &
 & $0.8$ & $0.6$ &$0.4$ & $0.2$\\  
  \hline
$500$ & $0.92$ & $0.91$ & $0.91$& $0.94$ &
&  $0.80$ & $0.90$ & $0.92$ & $0.95$ &
&  $0.82$ & $0.92$ & $0.93$ & $0.93$\\ 

  $1000$ & $0.96$ & $0.92$ & $0.94$ & $0.94$ &
 & $0.60$ & $0.87$ & $0.93$ & $0.94$ &
  & $0.82$ & $0.91$ & $0.94$ & $0.96$ \\ 
  
  $1500$ & $0.94$ & $0.94$ & $0.91$ & $0.94$ &
  & $0.39$ & $0.85$ & $0.91$ & $0.94$ &
   & $0.84$ & $0.90$ & $0.94$ & $0.94$\\ \hline

   \end{tabular*}
\end{table}

The confidence intervals constructed using $\hat{\tau}_{\textsc{AIPW}}^{\text{refit}} - b^*$  have minimum empirical coverage of $0.91.$
However, when a plug-in estimator of the confounding bias $\hat{b}^{\text{refit}}$ is used
low empirical coverage may arise when $\rho$ is not small enough (relative to the sample size). In particular, when model selection is performed, the coverages of the confidence intervals calculated using $\hat{\tau}_{\textsc{AIPW}}^{\text{refit}} - \hat{b}^{\text{refit}}$ are low if $\rho$ is large (relative to the sample size). The minimum coverage is 0.39, when  $\rho=0.8$. 
However, as expected from Theorem \ref{theorem:var_notcorrected_inference}, coverages get closer to the nominal level at smaller $\rho$ values for any given sample size.
Because of the weaker condition on $\rho$ in Theorem \ref{theorem:var_corrected_inference}, the coverage for the estimator $\hat{\tau}_{\textsc{AIPW}}^{\text{refit}} - \hat{b}^{\text{refit}}_{\text{c}}$ is, as expected, closer to nominal level compared to the estimator $\hat{\tau}_{\textsc{AIPW}}^{\text{refit}} - \hat{b}^{\text{refit}}$, and this improvement is more pronounced for large sample sizes.

\section{Case study: Effect of maternal smoking on birth weight}

We re-examine  a study that aims to assess the effect of smoking during pregnancy on birth weight \citep{almond2005costs,cattaneo2010efficient,scharfstein2021semiparametric}. The data comes from an open-access sample of 4996 individuals, 
 a sub-sample of approximately 500 000 singleton births in Pennsylvania between 1989 and 1991
(See \citealt{scharfstein2021semiparametric} and resources therein for more details on the data). 
 Weight at birth in grams is the outcome variable. The treated group consists of pregnant women who smoked during pregnancy, and the control group consists of women who did not smoke.
 
 We use the same set of covariates as in  \citet{scharfstein2021semiparametric}. As they argue, sensitivity analysis is necessary to account for potential unobserved confounders, such as genetic factors not observed.
The observed covariates that we consider are maternal data including numerical (age and the number of prenatal visits) and categorical variables (education $-$less than high school, high school, more than high school$-$ and birth order $-$one, two, larger than two)  and binary variables including white, hispanic, married, foreign and alcohol use. Higher order and interaction terms of the numerical variables of order up to three are also considered. Finally, all the second-order interactions with categorical variables are added, which gives a total of 80 terms.

The analysis in \citet{almond2005costs} estimates that smoking has an average effect of -203.2 grams on birth weight, determined by a regression model without taking into account the effect of unobserved confounding. In \citet{scharfstein2021semiparametric}, a semiparametric approach yields an estimate of -223 grams (95\% confidence interval [-274, -172]). However, after accounting for unobserved confounding in a sensitivity analysis, the latter study suggests that the average effect is not more extreme than -200 grams; see \eqref{priorassumptions} below for details on the clinical assumptions used.

In our analysis, we use the AIPW estimator of average causal effect, $E(Y(1))-E(Y(0))$. The results presented in this paper are directly applicable to $E(Y(0))$ as well, and, therefore, to $E(Y(1))-E(Y(0))$ under corresponding assumptions. In particular, there are now two sensitivity parameters through the sensitivity model of Assumption \ref{assum:sensitivity_model}, for both $Y(1)$ and $Y(0)$, denoted respectively $\rho_1$ and $\rho_0$. We use \texttt{subset = 'refit'} in function  \texttt{ui.causal} to exploit the built-in variable selection function (\texttt{ui.causal} included in the \texttt{R}-package \texttt{hdim.ui}). The resulting AIPW estimation gives an estimate of the average causal effect of -221 grams (95\% confidence interval: [-273, -168]).

\begin{figure*}[!t]%
\centering
\centering\includegraphics[width=0.85\textwidth]{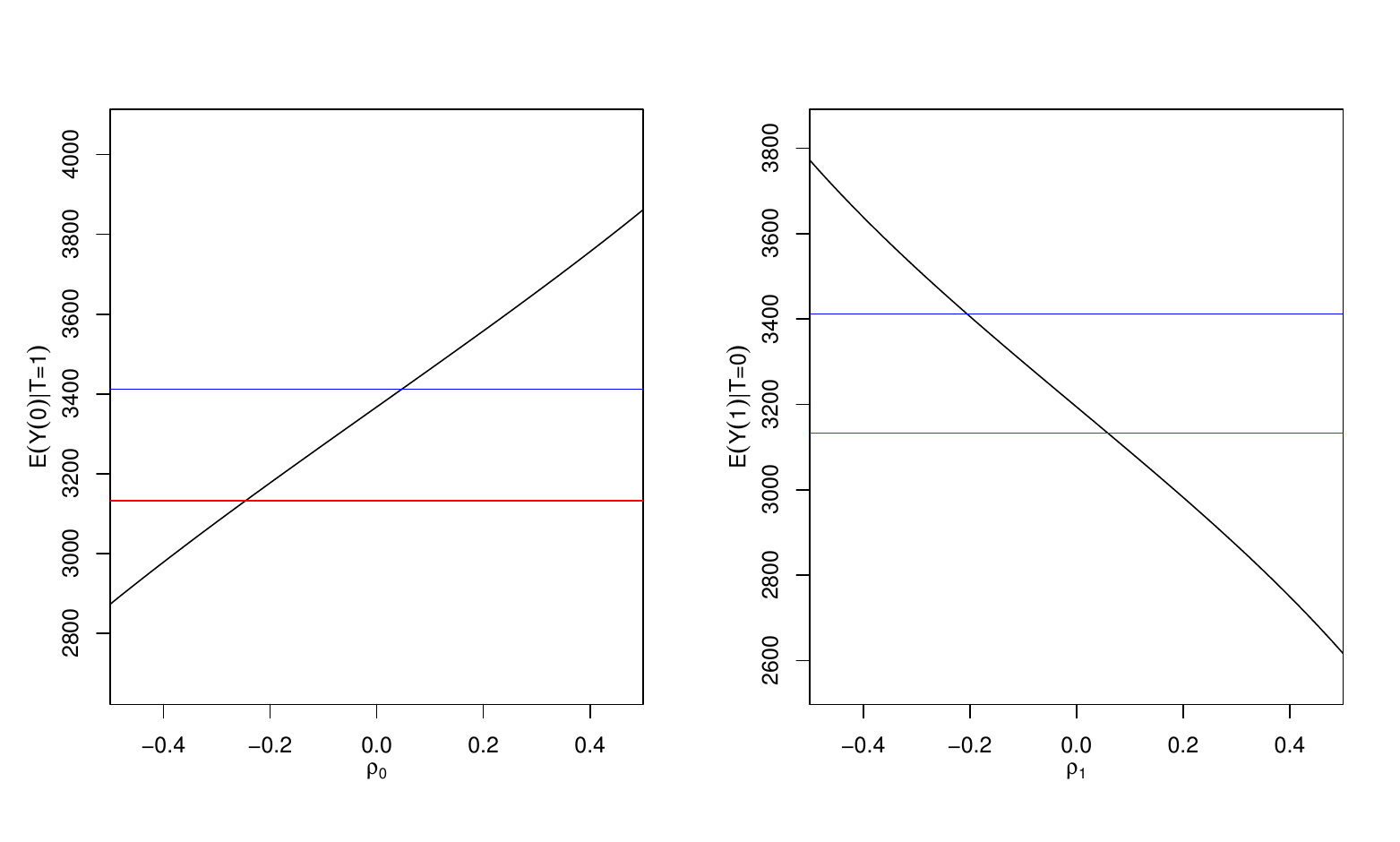}
   \caption{Left plot: estimated expection of birth weight for smokers under not smoking, $E(Y(0)|T=1)$. This estimate is constrained by the sample average estimations of $E(Y(1)|T=1)$ in red and $E(Y(0)|T=0)$ in blue, as\eqref{priorassumptions} suggests. Right plot: the estimation of $E(Y(1)|T=0)$. See Appendix for a description of the estimators. From these results, plausible values for the sensitivity parameters $\rho_0$ and $\rho_1$ are (-0.25, 0.05) and (-0.2, 0.06) respectively; i.e., where crossing with the red and blue bounds occur in respective plots.
  }\label{fig:rhobounds}
\end{figure*}

\begin{figure*}[!t]%
\centering
    \centering
   \includegraphics[width=0.85\textwidth]{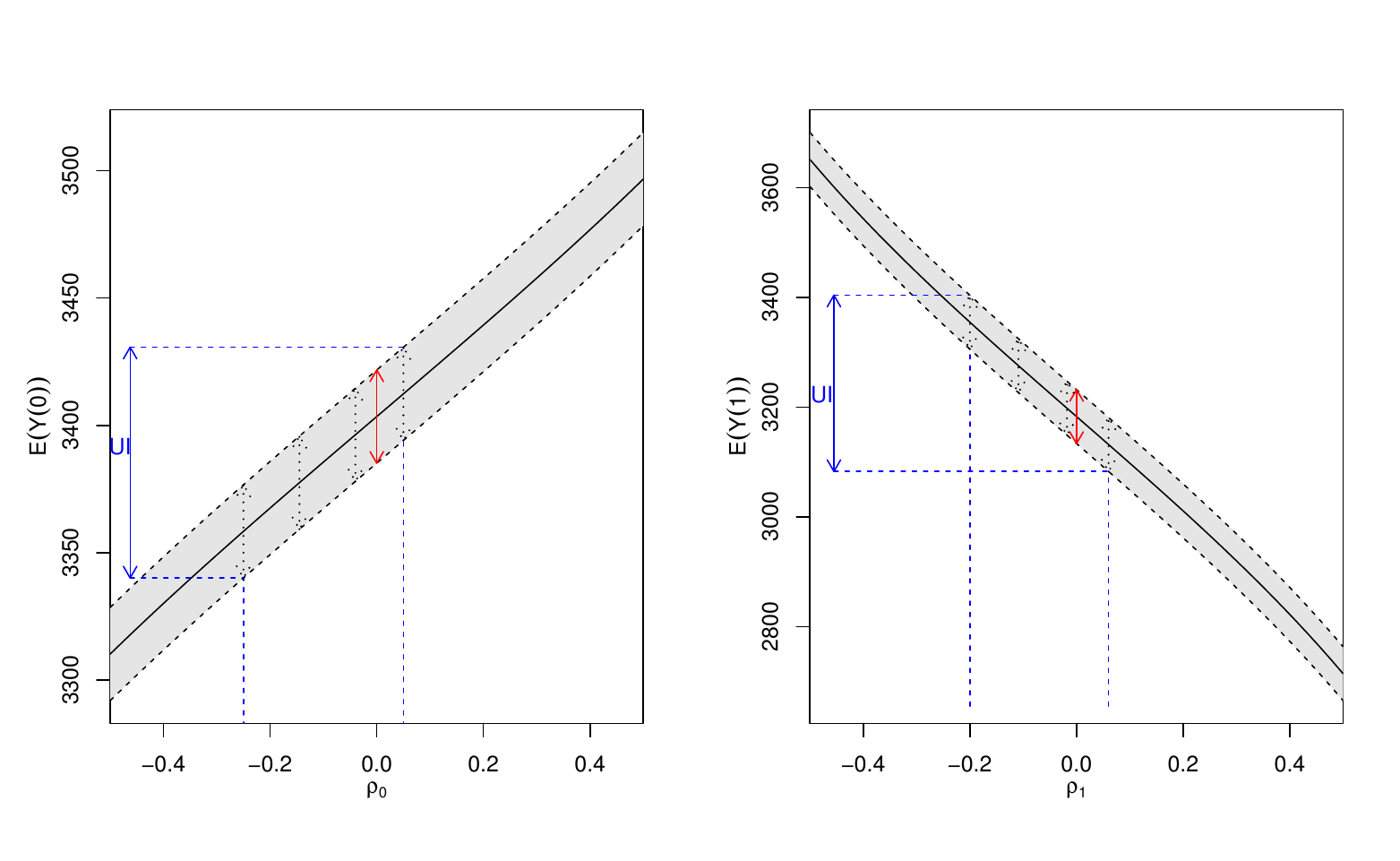}
    \caption{
    Estimates of average birth weight under smoking and non-smoking during pregnancy are given in the right and left plots, respectively; using output from \texttt{ui.causal} function (\texttt{hdim.ui} package). Solid black lines for bias-corrected AIPW estimates, $\hat{\tau}_{\textsc{AIPW}}^{\text{refit}} - \hat{b}^{\text{refit}}$,and 95\% confidence intervals (gray area). Red intervals are confidence intervals assuming no unobserved confounding ($\rho_t = 0, $ $t = 0,1$), and blue intervals are 95\% uncertainty intervals(using $-0.25 \leq \rho_0 \leq 0.05$ and $-0.20 \leq \rho_1 \leq 0.06$).
    }\label{fig:uiplots}
\end{figure*}

In their sensitivity analysis, \citet{scharfstein2021semiparametric} argued that the following inequalities make clinical sense:
\begin{equation}\label{priorassumptions}
    \begin{aligned}
    E(Y(1)|T=1)&<E(Y(0)|T=1)<E(Y(0)|T=0),\\
    E(Y(1)|T=1)&<E(Y(1)|T=0)<E(Y(0)|T=0).
    \end{aligned}
\end{equation}
For instance from the second line, under smoking, non-smokers are expected to have higher average birth weight than smokers, $E(Y(1)|T=0)>E(Y(1)|T=1)$. 
We choose to also use these prior clinical assumptions, which yield bounds for our sensitivity parameters $\rho_1$ and $\rho_0$. That is because the terms on the right and left-hand sides of the inequalities in \eqref{priorassumptions} can be unbiasedly estimated using sample averages since the data are fully observed, whereas the terms in the middle are functions of the sensitivity parameter. The Appendix provides a description of how they are estimated. According to the range of sensitivity parameters obtained from the above restrictions (See Figure \ref{fig:rhobounds}), we have derived estimates and the uniformly valid confidence intervals along with a resulting uncertainty interval for $E\left(Y\left(0\right)\right)$ and $E\left(Y\left(1\right)\right)$ (See Figure \ref{fig:uiplots}).
The resulting uncertainty interval for the average causal effect of maternal smoking on birth weight is [-333, 49].
Thus, while our AIPW estimation and 95\% confidence interval under the assumption that all confounders are observed are similar to those of \citet{scharfstein2021semiparametric}, the sensitivity analysis is quite different, since they suggest that the average effect is not more extreme than -200 grams while our analysis does not discard 
a potentially much larger adverse effect (up to 333 grams weight loss) of maternal smoking 
depending on the strength of confounding.

The difference in conclusions may be due to the fact that  \citet{scharfstein2021semiparametric} analysis excludes the unconfoundedness situation as one of possible scenarios while ours do not. This discrepancy might be due to different modeling assumptions and thereby different influence functions and corresponding estimation procedures. For instance, \citet{scharfstein2021semiparametric} sensitivity model establishes a link between the observed and unobserved potential outcome densities which necessitates the requirement of common supports; a condition stating that the support of each of the missing potential outcomes must be a subset of the support of the corresponding observed potential outcome.
This assumption may affect the validity of the results 
\citep[See section 7 in][]{franks2019flexible}.
 Furthermore, our method only uses clinical assumptions for bounding sensitivity parameters, whereas \citet{scharfstein2021semiparametric} uses those assumptions for both selecting a valid tilting function (defining the sensitivity model) and bounding sensitivity parameters. The effect on their analysis of using alternative tilting functions is unclear to us.

\section{Discussion}

Unobserved confounding cannot be discarded nor empirically investigated in observational studies and therefore sensitivity analysis to the unconfoundedness assumption should be common practice. Moreover, high-dimensional settings are typical in observational studies using large data sets and machine learning for nuisance models. We have presented here a novel method to conduct sensitivity analysis in such situations using uniformly valid estimators, which is essential in high-dimensional setting. In particular, the sensitivity analysis proposed is based on the construction of an uncertainty interval for the causal effect of interest which we show has uniformly valid coverage. Finite sample experiments confirm the asymptotic results. 

We use a sensitivity model with a sensitivity parameter which is easy to interpret and discuss with subject-matter scientists. As all sensitivity models, ours describes potential departures from the unconfoundedness assumption. If sensitivity is detected as is the case in the presented application on the effect of smoking on birth weight, then this is important information. If no sensitivity is detected then one might argue that this does not preclude the analysis to be sensitive to other departures from the unconfoundedness assumption.    
Lastly, note that our results show that we need to let $\rho$ tend to zero with increasing sample size unless we can estimate bias due to unobserved confounding, and hence the propensity score, with a root-n convergence rate. An alternative to bias estimation is to fix the bias itself as the sensitivity parameter, at the expense of interpretability.





\section{Appendix}

We use the notations $f(X)=E\left(Y\left(1\right)|X\right)$, $\Tilde{\sigma}^2 = E_{n_t}\left[\left(Y(1)_i - {m}(X_i)\right)^2\right]^{1/2}$and $v \bm{=} E((Y(1) - m(X))^2|T=1)$. Further, we use $a \leq_P b$ to denote $a=O_P(b)$.

\begin{appendixassumption}[Part of Assumption 2 in \citet{farrell2018robust}]\label{DGPfarrellassumptions} Let $U= Y(1) - m(X).$
$P_{n}$ obeys the following conditions, with bounds uniform in $n$.
\begin{itemize}
    \item[a.]  $E\left[|U|^{4} \mid X\right] \leq \mathcal{U}^{4}$.
    \item[b.] For some $r>0$ : $E\left[\left|m\left(x_{i}\right) 
    \right|^{2+2r}\right]$ and $E\left[\left|u_{i}\right|^{4+r}\right]$ are bounded.
    
\end{itemize}
\end{appendixassumption}

\begin{appendixassumption} \label{appendix.assum} We have $E\left(\lambda^4
    \left(\hat{g} \left(X\right)\right)\right) < \infty,$  $E({g}^4(X)) < \infty$, $E(g^2(X)\lambda^2(g(X))) < \infty$,  $E(\lambda^4({g}(X))) < \infty$, and $E(\lambda^2(\hat{g}(X) \lambda^2({g}(X))) <\infty$.
\end{appendixassumption}

 \begin{appendixassumption}\label{extraassumptions}Assume $E\left(\left(Y\left(1\right) - f\left(X\right)\right)^4\right) < \infty$, 
 $E \left(\lambda^2\left({g}\left(X\right)\right)\right)<\infty$,  
 \\$E\left(\left(Y(1) - \hat{m}(X)\right)^4\right) < \infty$, 
$E\left(g\left(X\right)\lambda\left(g\left(X\right)\right)|T=1\right)< \infty$  and
\newline
$E\left( \lambda^2\left(g\left(X\right)\right) |T=1)\right) <\infty$. 

\end{appendixassumption}

\begin{appendixassumption}\label{extraassumptionsalternatrive} Assume $E\left(\left(Y\left(1\right) - f\left(X\right)\right)^4\right) < \infty$,
$E\left(\left(Y(1) - \hat{m}(X)\right)^4\right) <~ \infty$,\\
$E({g}^2(X) \lambda^2({g}(X))) = O_P(1)$,
$E(\lambda^4({g}(X)))=O_P(1)$,
$\lambda(\hat{g}(X)) = O_P(1)$ and
${g}(X) = O_P(1)$.
\end{appendixassumption}

We frequently use the following lemma which is a direct result of \citet*[Theorem 14.1-1]{bishop2007discrete}. The lemma is used to translate some regularity conditions in the form of order in probability to moment conditions.
\begin{lemma}\label{bishoptheorem}
Assume that var$(A_{i})< C$ across $i$ and $n$, then $ 
E_n[A_{i}] - E(A_{i}) = O_P(n^{-1/2}).$
\end{lemma}

\subsection{Proof of Theorem \ref{theorem:asym_linearity_AIPW}} \label{app: proof_of_asymplinearitytheorem}

Suppose that Assumptions \ref{assum:overlap}, \ref{DGPfarrellassumptions}, and \ref{assum:rates} hold. By \citet[Theorem 3(1)]{farrell2015robust} we have
\begin{flalign*}
   &\sqrt{n}\left( E_n\left[\dfrac{T_i\left(Y_i-\hat{m}\left(X_i\right)\right)}{\hat{e}\left(X_i\right)} + \hat{m}\left(X_i\right)\right] - \tau^-  \right)\\&  = \dfrac{1}{\sqrt{n}} \sum_{i=1}^n \left( \dfrac{T_i\left(Y_i-{m}\left(X_i\right)\right)}{{e}(X_i)} + {m}(X_i) -  \tau^- \right) + o_P(1). &
\end{flalign*}

In the above representation, unlike \cite{farrell2015robust}, the expectation $m(X)=E\left(Y\left(1\right)\mid X,T=1\right)$ is not necessarily equal to $E\left(Y\left(1\right)\mid X \right)$ and, therefore, the asymptotic expectation of the \textsc{AIPW} estimator, denoted by $\tau^-,$ can be different from $\tau$. 
 
\subsection{Proof of Theorem \ref{theorem:asym_conf_bias_expression}}\label{proofofconfoundingbiastheorem}

The steps in the proof follow \citet{genback2019causal}. However, the parametric modelling assumptions are dropped here. By \citet[Theorem 2]{farrell2015robust} we have that under the consistency of one of the nuisance models and other regularity conditions specified $\hat{\tau}_\text{AIPW} \xrightarrow[]{p} E\left(m\left(X\right)\right)$. Also, 
\begin{flalign}\label{xiexpectation}
      & E(\xi\mid  X,T=1)  \notag\\&= E(\rho \sigma \eta + \epsilon \mid  X,T=1)
\notag\\&= \rho \sigma E(\eta\mid  \eta> -g(X) ) 
\notag\\&=\rho \sigma \lambda\left(g\left(X\right)\right).&
\end{flalign}

Using (\ref{xiexpectation}) and Assumption \ref{assum:sensitivity_model}, $
\tau^- = E\left( E\left(Y\left(1\right)\mid  X,T=1\right) \right) 
= \tau + \rho \sigma E\left( \lambda \left(g\left(X\right)\right) \right).
$
\subsection{Proof of Theorem \ref{theorem:var_notcorrected_inference}}\label{proof}
If we show that $\sqrt{n}(\hat{b} - b)=o_P(1)$, the asymptotic linearity result is a direct result of Theorem \ref{theorem:asym_linearity_AIPW}. We have
\begin{flalign*}
&\sqrt{n}(\hat{b} - b) \\&\begin{aligned}= & \sqrt{n}\rho  \hat{\sigma} E_n\left[\lambda\left(\hat{g}\left(X_i\right)\right) - \lambda\left({g}\left(X_i\right)\right) \right]\\&+ \sqrt{n}\rho  E_n\left[\lambda\left({g}\left(X_i\right)\right)\right] \left(\hat{\sigma} -  \sigma\right) \\& + \sqrt{n}\rho \sigma\left\{ E_n\left[\lambda\left({g}\left(X_i\right)\right)\right] - E\left(\lambda\left({g}\left(X_i\right)\right)\right)\right\}.\end{aligned}&
\end{flalign*}

For the first term, we have \begin{flalign*}
	\sqrt{n}\rho \hat{\sigma} E_n\left[\lambda\left(\hat{g}\left(X_i\right)\right) - \lambda\left({g}\left(X_i\right)\right) \right]  \leq_P   \sqrt{n} \rho E_n\left[\hat{g}\left(X_i\right) - g\left(X_i\right)\right]
	 = o_p (1),&\end{flalign*} where the inequality in probability is derived using Assumption \ref{extraassumptions}, Lemma \ref{bishoptheorem}, Lipschitz continuity of inverse Mills ratio while the  equality follows from Assumptions on $\rho$ in Theorem \ref{theorem:var_notcorrected_inference}. For the third term, using Lemma \ref{bishoptheorem} and Assumption \ref{extraassumptions}, one can show that	$	\sqrt{n}\rho\sigma\big\{ E_n\left[\lambda\left({g}\left(X_i\right)\right)\right] - E\left(\lambda\left({g}\left(X_i\right)\right)\right)\big\}= o_p (1)$. This shows that the first and third terms in the above decomposition are negligible. Therefore, using some moment conditions in Assumption \ref{extraassumptions}, we have 
	 \begin{flalign*}\label{bhatminusblimit}
	 &\sqrt{n}(\hat{b} - b) 
  \\&\leq_P \sqrt{n}\rho E_n\left[\lambda\left({g}\left(X_i\right)\right)\right] \left(\hat{\sigma} -  \sigma\right)
	\\& \leq_P  \sqrt{n}\rho  \left(\hat{\sigma}^2 -  \sigma^2\right)\\
   &\begin{aligned}\leq_P & \sqrt{n}\rho (\hat{\sigma}^2 - \Tilde{\sigma}^2)(=\bm{R_{21}}) \\
   &+ \sqrt{n}\rho (\Tilde{\sigma}^2 - v)(=\bm{R_{22}}) \\
& +  \sqrt{n}\rho \left( v - \sigma^2 \right) (=\bm{R_{23}}).\end{aligned}&
\end{flalign*}

For $\bm{R_{21}}$, using  Assumption \ref{extraassumptions} and a condition on $\rho$ stated 
we have 
\begin{flalign*}
& \sqrt{n}\rho (\hat{\sigma}^2 - \Tilde{\sigma}^2) \\ & \leq_P \sqrt{n}\rho (\hat{\sigma} - \Tilde{\sigma})\\
&\begin{aligned} \leq_P& \sqrt{n}\rho E_{n_t}\left[\left(Y(1)_i - {m}(X_i)\right)^2\right]^{1/2} \\
&+\sqrt{n}\rho E_{n_t}\left[\left( {m}(X_i)-\hat{m}\left(X_i\right)\right)^2\right]^{1/2}  \\
&- \sqrt{n}\rho E_{n_t}\left[\left(Y(1)_i - {m}(X_i)\right)^2\right]^{1/2} = o_P(1),\end{aligned}&
\end{flalign*}
where the second inequality in probability holds by Minkowski inequality.

For $\bm{R_{22}}$, using $\rho = o(1)$, Assumptions \ref{assum:overlap} and \ref{extraassumptions}
we have
\begin{flalign*}
&\sqrt{n}\rho (\Tilde{\sigma}^2 - v) \\
&\begin{aligned}=& 
\sqrt{n}\rho \left(\frac{n}{n_t}  E_{n} [T_i(Y(1)_i - {m}(X_i))^2] - v\right)\\
\leq_P& \sqrt{n}\rho \left(\frac{n}{n_t}  E_{n} [T_i(Y(1)_i - {m}(X_i))^2] - P(T=1)^{-1} E_{n} [T_i(Y(1)_i - {m}(X_i))^2]
\right)\\
&+ \sqrt{n}\rho \left(P(T=1)^{-1}  E_{n} [T_i(Y(1)_i - {m}(X_i))^2]  \right. \\&\left.  -P(T=1)^{-1} E\left(T\left(Y\left(1\right)  - {m}\left(X\right)\right)^2\right) \right)\\
\leq_P&\sqrt{n}\rho \left(E_n[T_i]^{-1}   - P(T=1)^{-1} \right)\\
&+ \sqrt{n}\rho \left( E_{n} [T_i(Y(1)_i - {m}(X_i))^2] -  E\left(T\left(Y\left(1\right) - {m}\left(X\right)\right)^2\right) \right)=o_P(1).\end{aligned}&
\end{flalign*}

For $\bm{R_{23}}$, we have
\begin{flalign*}
    &E((Y(1) - m(X))^2|T=1) \\
&\begin{aligned}
    =&E((Y(1) - f(X) - \rho \sigma \lambda(g(X)))^2|T=1)\\
    =& E((Y(1) - f(X))^2  + \rho^2 \sigma^2 \lambda^2(g(X)) -2 \rho \sigma (Y(1) - f(X)) \lambda(g(X)) |T=1) \\
    =& E((Y(1) - f(X))^2|T=1)+ \rho^2 \sigma^2 E( \lambda^2(g(X))|T=1)\\ &-2 \rho \sigma E((Y(1) - f(X)) \lambda(g(X)) |T=1),\end{aligned}&
\end{flalign*}
where 
\begin{flalign*}
   & E((Y(1) - f(X))^2|T=1) = -\sigma^2 \rho^2  E(g(X)\lambda(g(X))|T=1) + \sigma^2, &
\end{flalign*}
using Equation (A.2) in \cite{gorbach2018inference} and 
\begin{flalign*}
  &2 \rho \sigma E((Y(1) - f(X)) \lambda(g(X)) |T=1) = 2 \rho \sigma E(\xi \lambda(g(X)) |T=1)\\
  &=  2 \rho \sigma E((\rho\sigma\eta + \epsilon) \lambda(g(X)) |T=1) \\
 &=   2 \rho^2 \sigma^2 E(\eta \lambda(g(X)) |T=1) + 2 \rho \sigma E(\epsilon \lambda(g(X)) |T=1)\\
  &=   2 \rho^2 \sigma^2 E( \lambda(g(X))E(\eta|X,T=1) |T=1) + 2 \rho \sigma E( \lambda(g(X))E(\epsilon|X,T=1) |T=1)\\
  &= 2 \rho^2 \sigma^2 E( \lambda^2(g(X)) |T=1).&
\end{flalign*}
Therefore,
\begin{flalign*}
   & E((Y(1) - m(X))^2|T=1) \\&= \sigma^2  ( 1 - \rho^2E(g(X)\lambda(g(X))|T=1) - \rho^2  E( \lambda^2(g(X)) |T=1)).& \end{flalign*}

Finally, using assumptions on $\rho$ and Assumption \ref{extraassumptions}, we have the following for $\bm{R_{23}}$ 

\begin{flalign*}
&\sqrt{n}\rho \left( v - \sigma^2\right) \\&= \sqrt{n}\rho^3 E\left(g\left(X\right)\lambda\left(g\left(X\right)\right)|T=1\right) - \sqrt{n} \rho^3 \sigma^2 E\left( \lambda^2\left(g\left(X\right)\right) |T=1)\right)=o_p(1).&
\end{flalign*}

Theorem \ref{theorem:var_notcorrected_inference}(b) is a direct result of Theorem \ref{theorem:var_notcorrected_inference}(a) and the moment condition in Assumption \ref{assum:rates}(c).

Theorem \ref{theorem:var_notcorrected_inference}(c)
 holds based on the proof of Theorem 3.3 in \citet{farrell2015robust},
Assumptions \ref{assum:overlap}--\ref{assum:rates} and Theorem  \ref{theorem:var_notcorrected_inference}(b).
\subsection{Proof of Theorem 4}\label{consistencytheoremproof}

\textbf{Step 1. }First we find the limit of the term in the numerator of the variance estimator. By the triangle inequality, we have
\begin{flalign*}
 &| E_{n_t}^{1/2} [(Y(1)_i - {m}(X_i))^2] -  E_{n_t}^{1/2} [(m(X_i) - \hat{m}(X_i))^2]|\\&  \leq E_{n_t}^{1/2} [(Y(1)_i - \hat{m}(X_i))^2]\\ & \leq  E_{n_t}^{1/2} [(Y(1)_i - {m}(X_i))^2] +  E_{n_t}^{1/2} [(m(X_i) - \hat{m}(X_i))^2], &
\end{flalign*}
where using  Assumption \ref{assum:overlap} (which implies $n/n_t = O_P(1)$) and consistency of $\hat{m}(X)$ we have
\begin{flalign*}
E_{n_t}^{1/2} [(m(X_i) - \hat{m}(X_i))^2] & \leq \frac{n}{n_t}  E_{n}^{1/2} [T_i(m(X_i) - \hat{m}(X_i))^2] \xrightarrow[]{p} 0.&
\end{flalign*}

To  bound $E_{n_t}^{1/2} [(Y(1)_i - \hat{m}(X_i))^2]$ by the squeeze theorem, we just need to find the limit of $E_{n_t}^{1/2} [(Y(1)_i - {m}(X_i))^2]$. We have
\begin{flalign*}
 & E_{n_t}^{1/2} [(Y(1)_i - {m}(X_i))^2] 
   \\&= \frac{n}{n_t}  E_{n} [T_i(Y(1)_i - {m}(X_i))^2]\\
& \begin{aligned}\xrightarrow[]{p} &(P(T=1))^{-1} E((Y(1) - m(X))^2|T=1) P(T=1) \\&= E((Y(1) - m(X))^2|T=1),\end{aligned}&\end{flalign*}
where the convergence is the result of Lemma~\ref{bishoptheorem} and Assumption~\ref{appendix.assum}.

Note that based on the proof of Theorem \ref{theorem:var_notcorrected_inference} for $\bm{R_{23}}$, the limit found above has the following relationship with the true parameter $\sigma^2$. 
\begin{flalign*}
    &E((Y(1) - m(X))^2|T=1)\\& = \sigma^2  ( 1 - \rho^2E(g(X)\lambda(g(X))|T=1) - \rho^2  E( \lambda^2(g(X)) |T=1)). &
\end{flalign*}

\textbf{Step 2. }It now remains to show that
\begin{flalign*}
&E_{n_t}[\hat{g}(X_i) \lambda(\hat{g}(X_i))] - E_{n_t}[ \lambda^2(\hat{g}(X_i))]\\& \xrightarrow[]{p}  E(g(X)\lambda(g(X))|T=1) - E( \lambda^2(g(X)) |T=1).&
\end{flalign*}
We have 
\begin{flalign*}
&E_{n_t}[\hat{g}(X_i) \lambda(\hat{g}(X_i))]  - E(g(X)\lambda(g(X))|T=1) 
\\&\begin{aligned}=& \frac{n}{n_t} E_{n}[T\hat{g}(X_i) \lambda(\hat{g}(X_i))] - \frac{n}{n_t} E_{n}[T{g}(X_i) \lambda(\hat{g}(X))]\\ &+  \frac{n}{n_t} E_{n}[T{g}(X_i) \lambda(\hat{g}(X_i))]  - \frac{n}{n_t} E_{n}[T{g}(X_i) \lambda({g}(X_i))] \\ &+ \frac{n}{n_t} E_{n}[T{g}(X_i) \lambda({g}(X_i))] - E(g(X)\lambda(g(X))|T=1) \end{aligned}\\
&\begin{aligned}\leq& \frac{n}{n_t} E_{n}^{1/2}[\lambda^2(\hat{g}(X_i))] E_{n}^{1/2}[T(\hat{g}(X_i) - {g}(X_i))^2]\\ 
&+  \frac{n}{n_t}  E_{n}^{1/2}[{g}^2(X_i)] E_{n}^{1/2}[T( \lambda(\hat{g}(X_i))- \lambda({g}(X_i)))^2] \\ &+ \frac{n}{n_t} E_{n}[T{g}(X_i) \lambda({g}(X_i))] - E(g(X)\lambda(g(X))|T=1) \xrightarrow[]{p}  0,\end{aligned}&
\end{flalign*}
where the inequality is due to the Cauchy-Schwarz inequality and the convergences follows from Assumption \ref{assum:overlap},
 consistency assumption for $\hat{g}(X)$, Lipschitz continuity of the inverse Mills ratio $\lambda(.)$, Lemma~\ref{bishoptheorem} and Assumption~\ref{appendix.assum}.
Similarly,  we have
\begin{flalign*}
&E_{n_t}[ \lambda^2(\hat{g}(X_i))] - E( \lambda^2(g(X)) |T=1)
\\& =  \frac{n}{n_t} \left(E_{n}[ \lambda^2(\hat{g}(X_i))] -  E_{n}[ \lambda^2({g}(X_i))]  + E_{n}[ \lambda^2({g}(X_i))]\right) - E( \lambda^2(g(X)) |T=1) \\
 &\begin{aligned}= & \frac{n}{n_t} E_{n}[ (\lambda(\hat{g}(X_i)) - \lambda({g}(X_i))) (\lambda(\hat{g}(X_i)) + \lambda({g}(X_i)))] 
 \\ & + \frac{n}{n_t} E_{n}[ \lambda^2({g}(X_i))] - E( \lambda^2(g(X)) |T=1)\end{aligned} \\
 &\begin{aligned}\leq 
 & \frac{n}{n_t} E_{n}^{1/2}[ (\lambda(\hat{g}(X_i)) - \lambda({g}(X_i)))^2]E_{n}^{1/2}[ (\lambda(\hat{g}(X_i)) + \lambda({g}(X_i)))^2] 
 \\ & + \frac{n}{n_t} E_{n}[ \lambda^2({g}(X_i))] - E( \lambda^2(g(X)) |T=1) 
\xrightarrow[]{p} 0.\end{aligned}&
\end{flalign*}

\textbf{Step 3. }
The consistency of the causal parameter estimator can be shown by the consistency of the variance estimator $\hat{\sigma}^{2}_{\text{c}}$, Slutsky's theorem and $E_{n}[ \lambda(\hat{g}(X_i))] \xrightarrow[]{p} E( \lambda({g}(X_i)))$. The latter holds based on consistency of $\hat{g}(X)$, Lipschitz continuity of $\lambda(.)$, Lemma~\ref{bishoptheorem} and Assumption~\ref{appendix.assum}.

\subsection{Proof of Theorem \ref{theorem:var_corrected_inference}}
We have
\begin{flalign*}
 &\sqrt{n}(\hat{b} - b) \\&\leq_P \sqrt{n}\rho E_n\left[\lambda\left({g}\left(X_i\right)\right)\right] \left(\hat{\sigma}_\text{c} -  \sigma\right) 
  \leq_P\sqrt{n} \rho \left(\hat{\sigma}_\text{c}^2 - \sigma^2\right)\\
 &\begin{aligned}\leq_P& \sqrt{n}\rho (\hat{\sigma}^2 - \Tilde{\sigma}^2)/(1 - \rho^2 E_{n_t}[\hat{g}(X_i) \lambda(\hat{g}(X_i))] - \rho^2 E_{n_t}[ \lambda^2(\hat{g}(X_i))])(=\bm{R_{21}}) \\
   & + \sqrt{n}\rho (\Tilde{\sigma}^2 - v)/(1 - \rho^2 E_{n_t}[\hat{g}(X_i) \lambda(\hat{g}(X_i))] - \rho^2 E_{n_t}[ \lambda^2(\hat{g}(X_i))])(=\bm{R_{22}}) \\
& +  \sqrt{n}\rho \left( v/(1 - \rho^2 E_{n_t}[\hat{g}(X_i) \lambda(\hat{g}(X_i))] - \rho^2 E_{n_t}[ \lambda^2(\hat{g}(X_i))]) - \sigma^2 \right) (=\bm{R_{23}}). \end{aligned}&
\end{flalign*}
Where the first inequality can be shown from the first lines of the proof of Theorem \ref{theorem:var_notcorrected_inference} in Section \ref{proof}. Moreover, both $\bm{R_{21}}$ and $\bm{R_{22}}$ can be shown to be $o_p(1)$ from the proof of Theorem \ref{theorem:var_correction_consistency} regarding the terms with the same name and under some extra moment conditions in Assumption \ref{extraassumptionsalternatrive}. 
From the proof concerning the term named $\bm{R_{23}}$ in Theorem \ref{theorem:var_notcorrected_inference} we have the following. If Assumption \ref{assum:sensitivity_model} holds, we have
$v/(1 - \rho^2 E({g}(X_i) \lambda({g}(X_i))) - \rho^2 E(\lambda^2({g}(X_i)))) = \sigma^2.$
To complete the proof we have to show that 
\begin{flalign*}
 &\begin{aligned}\sqrt{n}\rho \big(& v/(1 - \rho^2 E_{n_t}[\hat{g}(X_i) \lambda(\hat{g}(X_i))]- \rho^2 E_{n_t}[ \lambda^2(\hat{g}(X_i))])  \\
& - v/(1 - \rho^2 E({g}(X_i) \lambda({g}(X_i))) - \rho^2 E(\lambda^2({g}(X_i))) \big)\end{aligned} \\&= o_p(1).    &
\end{flalign*}
Using Assumption \ref{extraassumptionsalternatrive} and assumptions on $\rho$ stated in the theorem we have
 \begin{flalign*}
&\bm{R_{23}}\\&\begin{aligned}
   \leq_P& \sqrt{n}\rho^3 \left( E_{n_t}[\hat{g}(X_i) \lambda(\hat{g}(X_i))]+E_{n_t}[ \lambda^2(\hat{g}(X_i))]\right) \\&- \sqrt{n}\rho^3 \left( E({g}(X_i) \lambda({g}(X_i))) -  E(\lambda^2({g}(X_i)) \right)\\ \leq_P&
   \sqrt{n}\rho^3 E_{n_t}[\hat{g}(X_i) \lambda (\hat{g}(X_i)) - {g}(X_i) \lambda (\hat{g}(X_i))]\\
   &+ \sqrt{n}\rho^3 E_{n_t}[{g}(X_i) \lambda (\hat{g}(X_i))- {g}(X_i) \lambda ({g}(X_i)) ]\\
   &+ \sqrt{n} \rho^3 E_{n_t}[ \lambda^2(\hat{g}(X_i)) -  \lambda^2({g}(X_i))] \\
   &+ \sqrt{n}\rho^3 \left( E_{n_t}[{g}(X_i) \lambda({g}(X_i))] -  E({g}(X_i) \lambda({g}(X_i)))\right) \\
   &+ \sqrt{n} \rho^3\left( E_{n_t}[ \lambda^2({g}(X_i))]  -E(\lambda^2({g}(X_i))  \right)
   = o_P(1),\end{aligned}&
\end{flalign*}
which completes the proof.

\subsection{Other parameters of interest} \label{otherparameters} Theorem~\ref{theorem:asym_linearity_AIPW} in the paper only concerns the causal parameter $E\left(Y\left(1\right)\right)$.
    The asymptotic linearity of the AIPW estimators of the causal parameters $E\left(Y\left(0\right)\right)$, $E\left(Y\left(1\right)|T=0\right)$ and $E\left(Y\left(0\right)|T=1\right)$ is shown under similar regularity conditions \citet[Theorem 3-4]{farrell2018robust}. Under asymptotic normality and using Assumption~\ref{assum:sensitivity_model},
    the asymptotic confounding bias of these estimators can be found similar to the one found for $E\left(Y\left(1\right)\right)$ in Theorem~\ref{theorem:asym_conf_bias_expression}. Here we show the proof for the parameter $\tau_{10} = E\left(Y\left(1\right)|T=0\right)$. We have
\begin{flalign*}
 &\begin{aligned}\dfrac{1}{P(T=0)}\bigg[&E\bigg(E((1-T)m(X)|X)\bigg)\\&+ E\bigg( E\big( \dfrac{T(Y-m(X)(1-e(X))}{e(X)}\vert X\big) \bigg)\bigg] - \tau_{10}\end{aligned}\\
  &\begin{aligned}=
\dfrac{1}{P(T=0)}\bigg[& E\big((1-e(X))m(X)\big)\\&+ E\big(\dfrac{1-e(X)}{{e}(X)}  E(T\big(Y-m(X)\big)\vert X) \big)\bigg]  - \tau_{10}\end{aligned}\\
&\begin{aligned}=
 \dfrac{1}{P(T=0)}\bigg[&E((1-e(X))m(X))\\&+E\big(\dfrac{1-e(X)}{{e}(X)} E(Y(1) - m(X)\vert X,T=1)Pr(T=1\vert X) \big)\bigg] - \tau_{10}\end{aligned}\\
 &=
 \dfrac{1}{P(T=0)}\bigg[E\bigg( (1-Pr(T=1\vert X)) E\big(Y(1)\vert X,T=1\big) \bigg)\bigg] - \tau_{10}\\
 &\begin{aligned}=
 \dfrac{1}{P(T=0)}\bigg[&E\bigg(  E\big(Y(1)\vert X,T=1\big) \\&-Pr(T=1\vert X) E\big(Y(1)\vert X,T=1\big) \bigg)\bigg]- \tau_{10}\end{aligned}\\
 &\begin{aligned}=
 \dfrac{1}{P(T=0)}\bigg[&E\bigg(  E\big(Y(1)\vert X\big) \\& + \rho \sigma \lambda\big(g(X)\big)  -Pr(T=1\vert X) E\big(Y(1)\vert X,T=1\big) \bigg)\bigg] - \tau_{10}\end{aligned}\\
 &\begin{aligned}=
   \dfrac{1}{P(T=0)}\bigg[ &\rho \sigma E\bigg(\lambda\big(g(X)\big)\\&+ Pr(T=0\vert X) E\big(Y(1)\vert X,T=0\big) \bigg) \bigg] - \tau_{10}\end{aligned}\\
   &=
  \dfrac{\rho \sigma E\bigg(\lambda\big(g(X)\big)\bigg)}{P(T=0)} + \dfrac{E\bigg(Pr(T=0\vert X) E\big(Y(1)\vert X,T=0\big)\bigg)}{P(T=0)}  - \tau_{10}\\
  &=
  \dfrac{\rho \sigma E\bigg(\lambda\big(g(X)\big)\bigg)}{P(T=0)} + \dfrac{E\bigg( E\big((1-T) Y(1)\vert X\big)\bigg)}{P(T=0)}  - \tau_{10}\\
&=  \dfrac{\rho \sigma E\bigg(\lambda\big(g(X)\big)\bigg)}{P(T=0)} + \dfrac{ E\big((1-T) Y(1)\big)}{P(T=0)}  - \tau_{10}\\
&=
  \dfrac{\rho \sigma E\bigg(\lambda\big(g(X)\big)\bigg)}{P(T=0)} + \dfrac{ P(T=0) E\big( Y(1)|T=0\big)}{P(T=0)}  - \tau_{10}
 \dfrac{\rho \sigma E\bigg(\lambda\big(g(X)\big)\bigg)}{P(T=0)}. &&
\end{flalign*}
For estimating the confounding bias at a given $\rho$ value we use $\rho \hat{\sigma}_{\text{c}} E_n[\lambda(\hat{g}(X_i))]/E_n[T_i]$.


\section{Supplementary material} Supplementary material contains additional simulation results and is available online at \newline\url{https://github.com/stat4reg/hdim.ui#readme}.

\section*{Acknowledgements} We are grateful to Minna Genbäck and Mohammad Ghasempour for their helpful comments. Funding from the Marianne and Marcus Wallenberg Foundation and the Swedish Research Council for Health, Working Life and Welfare is acknowledged.

  \bibliographystyle{chicago}
 \bibliography{bib.bib}

\begin{thebibliography}{}

\bibitem[\protect\citeauthoryear{Almond, Chay, and Lee}{Almond
  et~al.}{2005}]{almond2005costs}
Almond, D., K.~Y. Chay, and D.~S. Lee (2005).
\newblock The costs of low birth weight.
\newblock {\em The Quarterly Journal of Economics\/}~{\em 120\/}(3),
  1031--1083.

\bibitem[\protect\citeauthoryear{Belloni, Chernozhukov, and Hansen}{Belloni
  et~al.}{2014}]{belloni2014inference}
Belloni, A., V.~Chernozhukov, and C.~Hansen (2014).
\newblock Inference on treatment effects after selection among high-dimensional
  controls.
\newblock {\em The Review of Economic Studies\/}~{\em 81\/}(2), 608--650.

\bibitem[\protect\citeauthoryear{Bishop, Fienberg, and Holland}{Bishop
  et~al.}{2007}]{bishop2007discrete}
Bishop, Y.~M., S.~E. Fienberg, and P.~W. Holland (2007).
\newblock {\em Discrete multivariate analysis: theory and practice}.
\newblock Springer Science \& Business Media.

\bibitem[\protect\citeauthoryear{Bonvini and Kennedy}{Bonvini and
  Kennedy}{2022}]{bonvini2019sensitivity}
Bonvini, M. and E.~H. Kennedy (2022).
\newblock Sensitivity analysis via the proportion of unmeasured confounding.
\newblock {\em Journal of the American Statistical Association\/}~{\em
  117\/}(539), 1540--1550.

\bibitem[\protect\citeauthoryear{Cattaneo}{Cattaneo}{2010}]{cattaneo2010efficient}
Cattaneo, M.~D. (2010).
\newblock Efficient semiparametric estimation of multi-valued treatment effects
  under ignorability.
\newblock {\em Journal of Econometrics\/}~{\em 155\/}(2), 138--154.

\bibitem[\protect\citeauthoryear{Chernozhukov, Chetverikov, Demirer, Duflo,
  Hansen, Newey, and Robins}{Chernozhukov
  et~al.}{2018}]{chernozhukov2018double}
Chernozhukov, V., D.~Chetverikov, M.~Demirer, E.~Duflo, C.~Hansen, W.~Newey,
  and J.~Robins (2018).
\newblock {Double/debiased machine learning for treatment and structural
  parameters}.
\newblock {\em The Econometrics Journal\/}~{\em 21\/}(1), C1--C68.

\bibitem[\protect\citeauthoryear{Chernozhukov, Hansen, and
  Spindler}{Chernozhukov et~al.}{2016}]{chernozhukov2016hdm}
Chernozhukov, V., C.~Hansen, and M.~Spindler (2016).
\newblock hdm: High-dimensional metrics.
\newblock {\em arXiv preprint arXiv:1608.00354\/}.

\bibitem[\protect\citeauthoryear{Copas and Li}{Copas and
  Li}{1997}]{copas1997inference}
Copas, J.~B. and H.~G. Li (1997).
\newblock Inference for non-random samples.
\newblock {\em Journal of the Royal Statistical Society: Series B (Statistical
  Methodology)\/}~{\em 59\/}(1), 55--95.

\bibitem[\protect\citeauthoryear{Cornfield, Haenszel, Hammond, Lilienfeld,
  Shimkin, and Wynder}{Cornfield et~al.}{1959}]{cornfield1959smoking}
Cornfield, J., W.~Haenszel, E.~C. Hammond, A.~M. Lilienfeld, M.~B. Shimkin, and
  E.~L. Wynder (1959).
\newblock {Smoking and Lung Cancer: Recent Evidence and a Discussion of Some
  Questions}.
\newblock {\em JNCI: Journal of the National Cancer Institute\/}~{\em 22\/}(1),
  173--203.

\bibitem[\protect\citeauthoryear{Ding and VanderWeele}{Ding and
  VanderWeele}{2016}]{ding2016sensitivity}
Ding, P. and T.~J. VanderWeele (2016).
\newblock Sensitivity analysis without assumptions.
\newblock {\em Epidemiology (Cambridge, Mass.)\/}~{\em 27\/}(3), 368--377.

\bibitem[\protect\citeauthoryear{Farrell}{Farrell}{2015}]{farrell2015robust}
Farrell, M.~H. (2015).
\newblock Robust inference on average treatment effects with possibly more
  covariates than observations.
\newblock {\em Journal of Econometrics\/}~{\em 189\/}(1), 1--23.

\bibitem[\protect\citeauthoryear{Farrell}{Farrell}{2018}]{farrell2018robust}
Farrell, M.~H. (2018).
\newblock Robust inference on average treatment effects with possibly more
  covariates than observations.
\newblock {\em arXiv:1309.4686v3\/}.

\bibitem[\protect\citeauthoryear{Fisher}{Fisher}{1958}]{fisher1958cigarettes}
Fisher, R. (1958).
\newblock Cigarettes, cancer, and statistics.
\newblock {\em The Centennial Review of Arts \& Science\/}~{\em 2}, 151--166.

\bibitem[\protect\citeauthoryear{Franks, D’Amour, and Feller}{Franks
  et~al.}{2020}]{franks2019flexible}
Franks, A., A.~D’Amour, and A.~Feller (2020).
\newblock Flexible sensitivity analysis for observational studies without
  observable implications.
\newblock {\em Journal of the American Statistical Association\/}~{\em
  115\/}(532), 1730--1746.

\bibitem[\protect\citeauthoryear{Friedman, Hastie, Tibshirani, and
  Narasimhan}{Friedman et~al.}{2021}]{friedman2021package}
Friedman, J., T.~Hastie, R.~Tibshirani, and B.~Narasimhan (2021).
\newblock Package ‘glmnet’.
\newblock {\em CRAN R Repositary\/}.

\bibitem[\protect\citeauthoryear{Gabriel, Sj{\"o}lander, and Sachs}{Gabriel
  et~al.}{2023}]{erin:2023}
Gabriel, E.~E., A.~Sj{\"o}lander, and M.~C. Sachs (2023).
\newblock Nonparametric bounds for causal effects in imperfect randomized
  experiments.
\newblock {\em Journal of the American Statistical Association\/}~{\em
  118\/}(541), 684--692.

\bibitem[\protect\citeauthoryear{Genb\"ack and de~Luna}{Genb\"ack and
  de~Luna}{2019}]{genback2019causal}
Genb\"ack, M. and X.~de~Luna (2019).
\newblock Causal inference accounting for unobserved confounding after outcome
  regression and doubly robust estimation.
\newblock {\em Biometrics\/}~{\em 75\/}(2), 506--515.

\bibitem[\protect\citeauthoryear{Gorbach and de~Luna}{Gorbach and
  de~Luna}{2018}]{gorbach2018inference}
Gorbach, T. and X.~de~Luna (2018).
\newblock Inference for partial correlation when data are missing not at
  random.
\newblock {\em Statistics \& Probability Letters\/}~{\em 141}, 82--89.

\bibitem[\protect\citeauthoryear{Hines, Dukes, Diaz-Ordaz, and
  Vansteelandt}{Hines et~al.}{2022}]{hines2021demystifying}
Hines, O., O.~Dukes, K.~Diaz-Ordaz, and S.~Vansteelandt (2022).
\newblock Demystifying statistical learning based on efficient influence
  functions.
\newblock {\em The American Statistician\/}~{\em 76\/}(3), 292--304.

\bibitem[\protect\citeauthoryear{Imai, Keele, and Tingley}{Imai
  et~al.}{2010}]{imai2010general}
Imai, K., L.~Keele, and D.~Tingley (2010).
\newblock A general approach to causal mediation analysis.
\newblock {\em Psychological methods\/}~{\em 15\/}(4), 309--334.

\bibitem[\protect\citeauthoryear{Moosavi, H{\"a}ggstr{\"o}m, and
  de~Luna}{Moosavi et~al.}{2023}]{moosavi2021costs}
Moosavi, N., J.~H{\"a}ggstr{\"o}m, and X.~de~Luna (2023).
\newblock {The Costs and Benefits of Uniformly Valid Causal Inference with
  High-Dimensional Nuisance Parameters}.
\newblock {\em Statistical Science\/}~{\em 38\/}(1), 1 -- 12.

\bibitem[\protect\citeauthoryear{{R Core Team}}{{R Core Team}}{2019}]{R}
{R Core Team} (2019).
\newblock {\em R: A Language and Environment for Statistical Computing}.
\newblock Vienna, Austria: R Foundation for Statistical Computing.

\bibitem[\protect\citeauthoryear{Robins, Rotnitzky, and Zhao}{Robins
  et~al.}{1994}]{robins1994estimation}
Robins, J.~M., A.~Rotnitzky, and L.~P. Zhao (1994).
\newblock Estimation of regression coefficients when some regressors are not
  always observed.
\newblock {\em Journal of the American Statistical Association\/}~{\em
  89\/}(427), 846--866.

\bibitem[\protect\citeauthoryear{Rosenbaum}{Rosenbaum}{1987}]{rosenbaum1987sensitivity}
Rosenbaum, P.~R. (1987).
\newblock {Sensitivity analysis for certain permutation inferences in matched
  observational studies}.
\newblock {\em Biometrika\/}~{\em 74\/}(1), 13--26.

\bibitem[\protect\citeauthoryear{Rubin}{Rubin}{1974}]{rubin1974estimating}
Rubin, D.~B. (1974).
\newblock Estimating causal effects of treatments in randomized and
  nonrandomized studies.
\newblock {\em Journal of educational Psychology\/}~{\em 66\/}(5), 688--701.

\bibitem[\protect\citeauthoryear{Rubin}{Rubin}{1990}]{DR:90}
Rubin, D.~B. (1990).
\newblock Formal mode of statistical inference for causal effects.
\newblock {\em Journal of Statistical Planning and Inference\/}~{\em 25\/}(3),
  279--292.

\bibitem[\protect\citeauthoryear{Scharfstein, Rotnitzky, and
  Robins}{Scharfstein et~al.}{1999}]{scharfstein1999rejoinder}
Scharfstein, D., A.~Rotnitzky, and J.~Robins (1999).
\newblock Rejoinder to comments on “adjusting for non-ignorable drop-out
  using semiparametric non-response models?”.
\newblock {\em Journal of the American Statistical Association\/}~{\em 94},
  1121--1146.

\bibitem[\protect\citeauthoryear{Scharfstein, Nabi, Kennedy, Huang, Bonvini,
  and Smid}{Scharfstein et~al.}{2021}]{scharfstein2021semiparametric}
Scharfstein, D.~O., R.~Nabi, E.~H. Kennedy, M.-Y. Huang, M.~Bonvini, and
  M.~Smid (2021).
\newblock Semiparametric sensitivity analysis: Unmeasured confounding in
  observational studies.
\newblock {\em arXiv preprint arXiv:2104.08300\/}.

\bibitem[\protect\citeauthoryear{Tibshirani}{Tibshirani}{1996}]{tibshirani1996regression}
Tibshirani, R. (1996).
\newblock Regression shrinkage and selection via the lasso.
\newblock {\em Journal of the Royal Statistical Society: Series B
  (Methodological)\/}~{\em 58\/}(1), 267--288.

\bibitem[\protect\citeauthoryear{{Van der Laan} and Gruber}{{Van der Laan} and
  Gruber}{2010}]{van2010collaborative}
{Van der Laan}, M.~J. and S.~Gruber (2010).
\newblock Collaborative double robust targeted maximum likelihood estimation.
\newblock {\em The International Journal of Biostatistics\/}~{\em 6\/}(1),
  Article 17.

\bibitem[\protect\citeauthoryear{Van~der Laan and Rose}{Van~der Laan and
  Rose}{2011}]{vdLR:11}
Van~der Laan, M.~J. and S.~Rose (2011).
\newblock {\em Targeted learning: causal inference for observational and
  experimental data}.
\newblock Springer Science \& Business Media.

\bibitem[\protect\citeauthoryear{Vansteelandt, Goetghebeur, Kenward, and
  Molenberghs}{Vansteelandt et~al.}{2006}]{vansteelandt2006ignorance}
Vansteelandt, S., E.~Goetghebeur, M.~G. Kenward, and G.~Molenberghs (2006).
\newblock Ignorance and uncertainty regions as inferential tools in a
  sensitivity analysis.
\newblock {\em Statistica Sinica\/}~{\em 16\/}(3), 953--979.

\bibitem[\protect\citeauthoryear{Zhang and {Tchetgen Tchetgen}}{Zhang and
  {Tchetgen Tchetgen}}{2022}]{zhang2019semiparametric}
Zhang, B. and E.~J. {Tchetgen Tchetgen} (2022).
\newblock {A semi‐parametric approach to model‐based sensitivity analysis
  in observational studies}.
\newblock {\em Journal of the Royal Statistical Society Series A\/}~{\em
  185\/}(S2), 668--691.

\bibitem[\protect\citeauthoryear{Zhao, Small, and Bhattacharya}{Zhao
  et~al.}{2019}]{zhao2019sensitivity}
Zhao, Q., D.~S. Small, and B.~B. Bhattacharya (2019).
\newblock {Sensitivity analysis for inverse probability weighting estimators
  via the percentile bootstrap}.
\newblock {\em Journal of the Royal Statistical Society Series B\/}~{\em
  81\/}(4), 735--761.

\end{thebibliography}
\end{document}